\documentclass[final,5p,times,twocolumn]{elsarticle}

\usepackage{color,graphicx}
\usepackage{subfigure}
\usepackage{amsmath,amssymb,ascmac}
\usepackage{lineno,hyperref}
\modulolinenumbers[5]

\journal{Journal of \LaTeX\ Templates}

\begin{document}

\begin{frontmatter}

\title{Effects of three-baryon forces on kaon condensation in hyperon-mixed matter}

\author[1]{Takumi Muto\corref{cor1}}
\ead{takumi.muto@it-chiba.ac.jp}

\author[2]{Toshiki Maruyama}
\ead{maruyama.toshiki@jaea.go.jp}

\author[3]{Toshitaka Tatsumi}
\ead{tatsumitoshitaka@gmail.com}

\cortext[cor1]{Corresponding author}

\address[1]{Department of Physics, Chiba Institute of Technology, 2-1-1 Shibazono, Narashino, Chiba 275-0023, Japan}
\address[2]{Advanced Science Research Center, Japan Atomic Energy Agency , Ibaraki 319-1195, Japan} 
\address[3]{Institute of Education, Osaka Sangyo University, 3-1-1Nakagaito, Daito, Osaka 574-8530, Japan}

\begin{abstract}
Possibility of kaon-condensed phase in hyperon-mixed matter is considered on the basis of chiral symmetry for kaon-baryon and kaon-kaon interactions, 
being combined with the relativistic mean-field theory for two-body baryon interaction. 
In addition, universal three-baryon repulsive force in the string-junction model and phenomenological three-nucleon attractive force are introduced.
It is shown that softening of the equation of state stemming from both kaon condensation and mixing of hyperons is compensated with the repulsive effect of the three-baryon force and the relativistic effect for two-body baryon-baryon interaction. The latter effect reflects the density-dependence of scalar and vector meson mean-fields, which is constrained by the contribution of the attractive three-nucleon force to the binding energy at saturation density. 
The kaon-condensed phase in hyperon-mixed matter becomes stiff enough to be consistent with recent observations of massive neutron stars. 
\end{abstract}


\end{frontmatter}

\section{Introduction}
 As a novel form with macroscopic appearance of strangeness in strongly interacting matter, kaon condensation has longly received much attention in the interdisciplinary fields of particle-nuclear physics, astrophysics, and condensed matter physics~\cite{kn86,t88,mt92,tpl94,kvk95,lbm95,tstw98,fmmt96}. 
Driving force of kaon condensation is the $s$-wave kaon ($K$)-nucleon ($N$) interaction specified by chiral symmetry~\cite{kn86,t88}. 
 Kaon condensation most likely occurs when the lowest $K^-$ energy, which decreases with baryon density due to the $s$-wave $K$-$N$ attraction, meets the electron chemical potential~\cite{mt92,tpl94}. It has been shown that the equation of state (EOS) is largely softened in dense matter, once kaon condensation sets in~\cite{tpl94,fmmt96}. 
It has also been suggested that mixing of hyperons ($Y$) occurs in the ground state of neutron star matter~\cite{g85,nyt02,burgio2021}. 
The existence of hyperons would make the EOS soft as well. These phases lead to more compact stars reducing the maximum mass of neutron star and its radius~\cite{fmmt96,nyt02,burgio2021}. They would also affect thermal evolution of neutron stars through extraordinary cooling processes via enhanced neutrino emissions~\cite{t88,bkpp88,fmtt94,plp1992,gvm2018}.  

Recent multi-messenger observations with radio waves, $X$-rays, and gravitational waves associated with neutron star phenomena have provided with important information on various phases and properties of highly dense matter: The observations of massive neutron stars as large as 2 $M_\odot$ ($M_\odot$ being the solar mass) have put stringent constraint on the EOS~\cite{demo10,ant13,c2020,romani2021}. 
The detection of gravitational waves from neutron-star mergers (GW170817)~\cite{abbott2018,horowitz2018} and 
measurements of mass and radius of neutron stars through $X$-ray observation by Neutron star Interior Composition ExploreR (NICER)~\cite{riley2019,miller2019} have shed light on constraining the EOS of dense matter.

In a series of our works, we have studied possible coexistent phase with kaon condensates (KC) and $Y$-mixed matter [abbreviated as ($Y+K$) phase]~\cite{m2008,mmtt2019,mmt2021-1}. 
It has been shown that both KC and $Y$-mixing lead to significant softening of the EOS as a consequence of the combined effects of decreasing energy by the $s$-wave $K$-baryon ($B$) attraction and avoiding the $N$-$N$ repulsion at high densities by $Y$-mixing~\cite{nyt02}.
Thus the maximum mass of neutron stars with the ($Y$+$K$) phase looks too low to be compatible with observations of massive neutron stars. 

In the case of pure hyperon-mixed matter, the problem originating from such dramatic softening of the EOS has been called ``hyperon puzzle''. 
To resolve the hyperon puzzle, the necessity of introducing the universal three-baryon repulsion (UTBR) among hyperons and nucleons ($YYY$, $YYN$, $YNN$) as well as three-nucleon ($NNN$) force was pointed out~\cite{nyt02}.
 Subsequently there appeared several works taking into account three-body and multi-body forces between baryons such as multi-pomeron exchange potential~\cite{yamamoto2014,yamamoto2017} and  
$B$-meson ($M$)$M$, $MMM$ type diagrams in the relativistic mean-field (RMF) models~\cite{to12}. 
The diffusion Monte Carlo study including $\Lambda N$ and $\Lambda NN$ interactions has been done~\cite{lonardoni2015}, and recently the chiral effective field theory has been applied for the $YNN$ interaction~\cite{kohno2018,lvb2019,gkw2020,petschauer2020}. 
There are some other models for resolving the hyperon puzzle, e.~g.~, the SU(3) symmetry model for vector meson-$Y$ couplings~\cite{wcs2012}, the model with scaling of hyperon masses and couplings~\cite{mkv2015}, etc. 

As for kaon condensation, importance of nuclear three-body force for onset density of KC and the EOS of kaon-condensed phase in neutron-star matter has been discussed~\cite{zuo2004,Li2006}. 
In the ($Y$+$K$) phase, it may also be legitimate to assume that three-body repulsions among baryons ($Y$ and $N$) at high densities should work on an equal footing as three-nucleon repulsion, although there is few empirical information on the UTBR.

In the present work, in order to circumvent the significant softening of the EOS in the case of the ($Y$+$K$) phase, we take into account the repulsion by the UTBR. 
In the ($Y+K$) phase, both bosonic (KC) and fermionic (hyperons) degrees of freedom take part in the realization of strangeness in the ground state of matter.  In particular, we clarify how the former is affected by the UTBR for stiffening the EOS, while the latter is directly affected by receiving the repulsion from the UTBR. 
 
 We adopt the RMF model for two-body $B$-$B$ interaction mediated by meson-exchange, discarding the nonlinear self-interacting $\sigma, \omega$, or $\omega-\rho$ meson-coupling potentials~\cite{g85}. We call this model a minimal RMF (MRMF) throughout this paper. We introduce the density-dependent effective two-body potentials for the UTBR, which has been derived from the string-junction model by Tamagaki~\cite{t2008} and originally applied to $Y$-mixed matter by Tamagaki, Takatsuka and Nishizaki\cite{tnt2008}. Together with the UTBR,  phenomenological three-nucleon attraction (TNA) is taken into account, and we construct the baryon interaction model that reproduces saturation properties of symmetric nuclear matter (SNM) and empirical values of incompressibility, symmetry energy, and its slope at the nuclear saturation density $\rho_0$ (= 0.16 fm$^{-3}$). 
Then we consider the ($Y$+$K$) phase based upon the effective chiral Lagrangian coupled with the present baryon interaction model (MRMF+UTBR+TNA).  Effects of the UTBR and TNA on the whole EOS with the ($Y$+$K$) phase are clarified.  
As implications for thermal evolution of neutron stars, rapid cooling mechanisms in the presence of the ($Y$+$K$) phase are briefly mentioned. 
\vspace{-0.5cm}~
\section{Kaon condensation on the basis of chiral symmetry}
\label{sec:kaon}  

The $s$-wave $K$-$B$ scalar and vector interactions relevant to kaon condensation are embodied in the effective chiral Lagrangian~\cite{kn86}. 
 The former is simulated by the ``$K$-baryon sigma terms'' which explicitly break chiral symmetry, i.~e., $\Sigma_{Kb}$$\equiv (m_u+m_s)/2\cdot\langle b|\bar u u + \bar s s |b\rangle$ with $\langle b |\bar q q | b\rangle$ being the quark content in the baryon $b$. 
 The value of the $\Sigma_{Kn}$ is taken to be (300$-$400) MeV as standard values~\cite{mt92,mmt2021-1}. $\Sigma_{Kn}$ is closely related to the $\pi N$ sigma term, $\Sigma_{\pi N}$, for which we adopt the phenomenological value, $\Sigma_{\pi N}$ = 45 MeV~\cite{gls1991}. 
 The upper value for the $\Sigma_{Kn}$ (= 400 MeV) is obtained at leading order in chiral perturbation theory so as to be consistent with the octet baryon mass splitting. In this case, one has a large strangeness content in the nucleon, $y_N\equiv 2\langle N|\bar s s|N\rangle / \langle N|\bar u u+\bar d d|N\rangle =0.44$. 
 The lower value corresponds to the case 
$y_N \simeq 0$~\cite{ohki08}. 
In this case, $\Sigma_{Kn}$ is related to $\Sigma_{\pi N}$ by $\Sigma_{Kn}=\Sigma_{\pi n}(m_u+m_s)/[2(m_u+m_d)]$ on the assumption $\langle n|\bar u u|n \rangle\sim \langle n|\bar d d|n \rangle$. With $\Sigma_{\pi n}$ = 45 MeV, $m_u$ = 6 MeV, $m_d$ = 12 MeV, and $m_s$ = 240 MeV~\cite{kn86}, one obtains $\Sigma_{Kn}\sim$ 300 MeV. (Recent result for the quark masses from the lattice QCD, $m_u$=2.2 MeV, $m_d$ = 4.7 MeV, and $m_s$ = 95 MeV, little changes the result. )
 The latter vector interaction, corresponding to the Tomozawa-Weinberg term for the meson-$N$ scattering amplitude, is proportional to the term: 
$ X_0\equiv\left(\rho_p+\frac{1}{2}\rho_n-\frac{1}{2}\rho_{\Sigma^-}-\rho_{\Xi^-} \right)/(2f^2)$, 
where each coefficient in front of the number density of baryon, $\rho_b$, is specified as the $V$-spin charge, and $f$ the meson decay constant, for which we simply take the pion decay constant ($\simeq$ 93 MeV) in lowest-order in chiral perturbation. 
The classical kaon field is represented as $K^\pm=(f/\sqrt{2})\theta\exp(\pm i\mu_K t) $, where $\theta$ is the chiral angle and $\mu_K$ the kaon chemical potential. The Lagrangian density for the classical kaon field reads~\cite{mmt2021-1}
\begin{equation}
{\cal L}_K=f^2\Big\lbrack\frac{1}{2}(\mu_K\sin\theta)^2 - m_K^2(1-\cos\theta)
+2 \mu_K X_0 (1-\cos\theta)\Big\rbrack \ ,
\label{eq:lagk}
\end{equation}
where, $m_K$ is the free kaon mass, and the last term in the bracket stands for the $s$-wave $K$-$B$ vector interaction. The $s$-wave $K$-$B$ scalar interaction is absorbed into the effective baryon mass $M_b^\ast$. 
It is to be noted that the $K$-$K$ nonlinear self-interaction is naturally incorporated through the terms proportional to $\sin^2\theta$ and $\cos \theta$ as a consequence of the nonlinear representation of the $K$-field in the effective chiral Lagrangian. 

\section{Baryon interactions}
\label{sec:Bforce}

\subsection{Minimal RMF for two-body baryon interaction}
\label{subsec:MRMF}

The Lagrangian density for baryons and mesons which describes the two-body interaction is given by
\begin{eqnarray}
\hspace{-1.0cm}~{\cal L}_{BM}&=&\sum_{b}\overline{\psi}_b \left(i\gamma^\mu D_\mu^{(b)}-\widetilde M_b^\ast \right) \psi_b 
+\frac{1}{2}\left(\partial^\mu\sigma\partial_\mu\sigma - m_\sigma^2\sigma^2\right) \cr
&+&\frac{1}{2}\left(\partial^\mu\sigma^\ast\partial_\mu\sigma^\ast-m_{\sigma^\ast}^2\sigma^{\ast 2}\right)
-\frac{1}{4}\omega^{\mu\nu}\omega_{\mu\nu}+\frac{1}{2}m_\omega^2\omega^\mu\omega_\mu \cr
&-&\frac{1}{4}R_a^{\mu\nu} R^a_{\mu\nu}+\frac{1}{2}m_\rho^2 R_a^\mu R^a_\mu
- \frac{1}{4}\phi^{\mu\nu}\phi_{\mu\nu}+\frac{1}{2}m_\phi^2\phi^\mu\phi_\mu \ ,
\label{eq:lagbm}
\end{eqnarray}
where $\psi_b$ stands for baryon field $b$, and $M_b^\ast$ is the effective baryon mass defined by  $M_b^\ast\equiv M_b-g_{\sigma b}\sigma-g_{\sigma^\ast b}\sigma^\ast-\Sigma_{Kb}(1-\cos\theta)$ with $M_b$ being the free bayon mass and $g_{\sigma b}$, $g_{\sigma^\ast b}$ being the scalar meson-baryon  coupling constants. The vector meson fields for the $\omega$, $\rho$, $\phi$ mesons
are denoted as $\omega^\mu$, $R_a^\mu$ with the isospin component $a$, and $\phi^\mu$ ($\sim\bar s\gamma^\mu s$), respectively. The kinetic terms of the vector mesons are given in terms of 
$\omega^{\mu\nu}\equiv \partial^\mu \omega^\nu-\partial^\nu \omega^\mu$, $R_a^{\mu\nu}\equiv \partial^\mu R_a^\nu-\partial^\nu R_a^\mu$, and $\phi^{\mu\nu}\equiv \partial^\mu \phi^\nu-\partial^\nu \phi^\mu$. The vector meson-baryon couplings are introduced through the covariant derivative, $D_\mu^{(b)}\equiv \partial_\mu+i g_{\omega b}\omega_\mu+i g_{\rho b} {\hat I}_3^{\ (b)} (R_3)_0 +ig_{\phi b}\phi_\mu$, where $g_{mb}$ is the vector meson-baryon coupling constant and ${\hat I}_3^{\ (b)}$ is a sign of the third component of the isospin for baryon $b$. 

The vector meson couplings for hyperons ($Y$) are here related with those for the nucleon $N$ by SU(6) symmetry~\cite{sdg94} as $g_{\omega\Lambda}=g_{\omega\Sigma^-}=2g_{\omega \Xi^-}=(2/3) g_{\omega N}$, $g_{\rho \Lambda}=0, g_{\rho\Sigma^-}=2g_{\rho\Xi^-}=2g_{\rho N} $, 
 $g_{\phi\Lambda}=g_{\phi\Sigma^-}=(1/2) g_{\phi\Xi^-}=-(\sqrt{2}/3) g_{\omega N}$.

The scalar ($\sigma$, $\sigma^\ast$) mesons-hyperon couplings are determined from the phenomenological analyses of recent hypernuclear experiments. 
The $\sigma$-$Y$ coupling constant, $g_{\sigma Y}$, is related with the potential depth of the hyperon $Y$ ($Y=\Lambda$, $\Sigma^-$, $\Xi^-$) at $\rho_B=\rho_0$ in SNM, $V_Y^N$, which is written in the RMF as  
\begin{equation}
V_Y^N=-g_{\sigma Y}\langle\sigma\rangle_0 +g_{\omega Y}\langle\omega_0\rangle_0 
+\partial{\cal E}_{\rm UTBR}/\partial \rho_Y \ ,
\label{eq:ypot}
\end{equation}
where $\langle\sigma\rangle_0$ and $\langle\omega_0\rangle_0$ are the meson mean fields at $\rho_B=\rho_0$ in SNM, and the last term comes from the energy density contribution from the UTBR, ${\cal E}_{\rm UTBR}$, which is derived from Eq.~(\ref{eq:aUTBR}) in Sec.~\ref{subsec:TBR}. By setting $V_\Lambda^N=-27$ MeV, $V_{\Sigma^-}^N$ = 23.5 MeV, 
and $V_{\Xi^-}^N$ = $-14$ MeV in Eq.~(\ref{eq:ypot})
~\cite{ghm16}, one obtains $g_{\sigma\Lambda}$, $g_{\sigma\Sigma^-}$, and $g_{\sigma\Xi^-}$. 
For the $\sigma^\ast$-$Y$ coupling constant, $g_{\sigma^\ast Y}$,  
we determined $g_{\sigma^\ast \Lambda}$ to be 7.2 so as to reproduce the empirical values of the separation energy $B_{\Lambda\Lambda}$($^{\ \ 11}_{\Lambda\Lambda}$Be), 
with use of the $B$-$B$ interaction model in the RMF extended to 
finite nuclei~\cite{mmt09,mmt14}.
Within our $B$-$B$ interaction model, $g_{\sigma^\ast\Xi^-}$ is taken to be 4.0, for which one obtains the theoretical values of the separation energies $B^{\rm th}_{\Xi}(^{ \ \ 15}_{\ \Xi(s)}$C) = 8.1 MeV and $B^{\rm th}_{\Xi}(^{\ \ 12}_{\ \Xi(s)}$Be) = 5.1 MeV, which are consistent with the empirical values deduced from the ``Kiso'' event, $\Xi^-$ + $^{14}$N $\rightarrow$ $^{15}_{\ \Xi}$C $\rightarrow$ $^{10}_{\ \Lambda}$Be + $^5_\Lambda$He~\cite{n15,hayakawa2021}.  The remaining unknown coupling constant, $g_{\sigma^\ast \Sigma^-}$, is simply set to be zero.

\subsection{Three-baryon repulsive force}
\label{subsec:TBR}

The effective two-body potential $U_{\rm SJM}$ is obtained from the three-body baryon interaction $W({\bf r}_1, {\bf r}_2, {\bf r}_3)$ in the string-junction model by integrating out variables of the third baryon  multiplying the short-range correlation (s.r.c.) function squared $f_{\rm src}^2({\bf r})$: 
\begin{equation}
U_{\rm SJM}(1,2; \rho_{\rm B}) =
\rho_{\rm B}\int d^3 {\bf r}_3W({\bf r}_1, {\bf r}_2;{\bf r}_3)f_{\rm src}^2({\bf r}_1-{\bf r}_3)f_{\rm src}^2({\bf r}_2-{\bf r}_3) 
\label{eq:usjm}
\end{equation}
with $W({\bf r}_1, {\bf r}_2; {\bf r}_3) = W_0 g({\bf r}_1-{\bf r}_3)g({\bf r}_2-{\bf r}_3)$, where $W_0$ ($\simeq$ 2 GeV) is the strength of the order of $B$-antibaryon ($\bar B$) excitation energy~\cite{t2008}, and $g({\bf r}_i-{\bf r}_j)$ is the wavefunction between $B_i$ and $B_j$. Taking the wavefunction $g({\bf r})$ as the Gaussian form, $g({\bf r})=\exp[-(r/\eta_c)^2]$ with $r=|{\bf r}_i-{\bf r}_j|$ and $\eta_c$ [= (0.45$\sim$ 0.50) fm] being the range of the repulsive core for baryon forces, one obtains
\begin{equation}
U_{\rm SJM}(r; \rho_{\rm B})=\frac{\rho_{\rm B}W_0}{2\pi^2}\int_0^\infty dq q^2 j_0(qr)\left(G_{\rm src}(q)\right)^2 \ , 
\label{eq:usjm3}
\end{equation}
where $G_{\rm src}(q)$ is the Fourier transform of the $B$-$B$ wavefunction with the s.r.c. and $j_0(qr)$ (=$\sin(qr)/(qr)$) is the spherical Bessel function.  
Here the approximate form of $U_{\rm SJM}$ is used as 
\begin{equation}
U_{\rm SJM2}(r; \rho_{\rm B}) = V_r \rho_{\rm B}(1+c_r\rho_{\rm B}/\rho_0)\exp[-(r/\lambda_r)^2)] \ ,
\label{eq:aUTBR}
\end{equation}
 where $V_r$=95 MeV$\cdot$fm$^3$, $c_r$=0.024, and $\lambda_r$=0.86 fm corresponding to $\eta_c$ = 0.50 fm for SJM2~\cite{tnt2008}. The $U_{\rm SJM}$ grows almost linearly with $\rho_{\rm B}$. 
Finally one obtains the effective two-body potential, 
$\widetilde U_{\rm SJM}(r:~\rho_{\rm B})~=f_{\rm src}(r) U_{\rm SJM}(r; \rho_{\rm B})$. 

\subsection{Three-nucleon attractive force}
\label{subsec:TNA}

As for the TNA, we adopt the density-dependent effective two-body potential by Nishizaki, Takatsuka and Hiura~\cite{nth1994}, which was phenomenologically introduced and the direct term of which agrees with the    expression by Lagaris and Pandharipande~\cite{lp1981} [we later call it LP~(1981)]: 
\begin{equation}
U_{\rm TNA}(r; \rho_{\rm B})=V_a\rho_{\rm B} \exp(-\eta_a \rho_{\rm B})\exp[-(r/\lambda_a)^2]
(\vec{\bf\tau}_1\cdot\vec{\bf\tau}_2)^2 \ ,
\label{eq:tna}
\end{equation}
where the range parameter $\lambda_a$ is fixed to be 2.0 fm. 
The $U_{\rm TNA}(r; \rho_{\rm B})$ depends upon not only density but also isospin $\tau_1\cdot\tau_2$. The parameters $V_a$ and $\eta_a$ are determined together with 
other parameters to reproduce the saturation properties of the SNM for the allowable values of $L$. 

\section{Energy density for the ($Y$+$K$) phase and saturation properties in SNM}
\label{sec:e}

The energy density for the ($Y$+$K$) phase is given as the sum of the KC, baryon and mesons for two-body baryon interaction, UTBR and TNA for three-body interaction, and leptons: ${\cal E}={\cal E}_K+{\cal E}_{B,M}+({\cal E}_{\rm UTBR}+{\cal E}_{\rm TNA})+{\cal E}_e$. 
The ground state energy for the ($Y+K$) phase is obtained under the charge neutrality condition 
and the $\beta$-equilibrium condition at a given density $\rho_{\rm B}$. 

\begin{figure}[h]~\vspace{-1.0cm}~
\begin{center}
\includegraphics[height=0.32\textwidth]{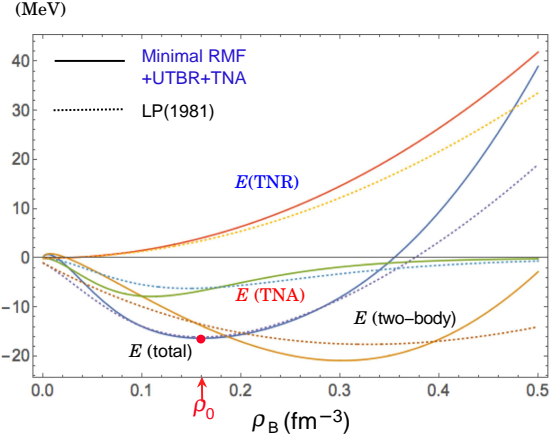}~
\end{center}~
\caption{The total energy per baryon, $E$ (total) $(={\cal E}/\rho_{\rm B}$), and each energy contribution to it from the three-nucleon-repulsion [$E$ (TNR)], the three-nucleon attraction [$E$(TNA)], and the sum of kinetic and two-body interaction energies [$E$(two-body)] in SNM are shown as functions of $\rho_{\rm B}$ by the solid lines in the case of $L$ = 65 MeV. For comparison, those obtained from LP~(1981)\cite{lp1981} are shown by the dotted lines. }
\label{fig1}
\end{figure}

The coupling constants, $g_{\sigma N}$, $g_{\omega N}$, and the meson mean-fields, $\langle\sigma\rangle_0$, $\langle\omega_0\rangle_0$ are determined so as to reproduce the properties of the SNM with saturation density $\rho_0$ and the binding energy $B_0$ (=16.3 MeV), together with the equations of motion for the $\sigma$ and $\omega$ mean-fields, $m_\sigma^2\langle\sigma\rangle_0=g_{\sigma N}\rho_N^s$ with the nuclear scalar density $\rho_N^s$, and $m_\omega^2\langle\omega_0\rangle_0=g_{\omega N}\rho_0$. Further, the coupling constant $g_{\rho N}$ and the parameters $\gamma_a$, $\eta_a$ in TNA associated with isospin-dependence are obtained to meet empirical values of the incompressibility $K$=240 MeV~\cite{GC2018}, the symmetry energy $S_0$ (=31.5 MeV)~\cite{lh2013} for a given value of the slope $L$, which is defined as  
$L$$\equiv 3\rho_0\left(\partial S / \partial \rho_B\right)_{\rho_B=\rho_0, x=1/2}=3 P_{\rm neutron \ matter}(\rho_0) / \rho_0 $. There is controversy about the empirical value of $L$, ranging from 30 MeV to 90 MeV~\cite{lp2016,oertel2017,xia2021}. 
For instance, there are several models with the smaller $L$ ($\lesssim$ 40 MeV), satisfying   
the constraints from compact star observations and HIC experiments~\cite{mkv2015,kmv2017}. 
On the other hand, there are rather larger values estimated from the  observations associated with $X$-ray bursters~\cite{sotani2015}. Recent PREX-II experiments on measurement of the neutron skin thickness of $^{208}$Pb have also reported a large value of $L$ = (73$-$146)~MeV~\cite{reed2021}. 
Further detailed analyses of experimental and observational information will be needed to 
constrain the precise value of the $L$. 
 We take the lower values $L$ = (60$-$70) MeV so that the density-dependence of the energy contributions around $\rho_0$ in SNM does not deviate much from those obtained by the standard variational calculation in terms of the $V_{14}$ two-body potential with addition of the phenomenological TNI by Lagaris-Pandahripande~[LP (1981)]~\cite{lp1981} [e.g., see Fig.~\ref{fig1} for  $L$ = 65 MeV].   
In Table~\ref{tab:para}, the relevant quantities with the MRMF+UTBR+TNA model are listed for typical cases of $L$ = (60, 65, 70) MeV. 

In Fig.~\ref{fig1}, the total energy per baryon, $E$~(total) $(={\cal E}/\rho_{\rm B}$), and each energy contribution from the three-nucleon-repulsion [$E$~(TNR)], the three-nucleon attraction [$E$~(TNA)], and the sum of kinetic and two-body interaction energies [$E$~(two-body)] in SNM are shown as functions of $\rho_{\rm B}$ obtained by the present model in the case of $L$ = 65 MeV by the solid lines. For comparison,  those of LP~(1981) are shown by the dotted lines, where the $E$~(total), $E$~(two-body) 
and $E$~(TNR) are read from Fig.~2, Tables~4 and 5 in~\cite{lp1981}, and the parameters in TNA in the case of LP (1981) are set to be $\gamma_a$ =$-$700 MeV$\cdot$fm$^6$ and $\eta_a$ = 13.6~fm$^3$~\cite{lp1981}. 

\begin{table*}[h]
\caption{The coupling constants, $g_{\sigma N}$, $g_{\omega N}$, $g_{\rho N}$, and the meson mean-fields, $\langle\sigma\rangle_0$, $\langle\omega_0\rangle_0$ in SNM in case of the MRMF+ UTBR+TNA model and the parameters $\gamma_a$, $\eta_a$ for TNA in case of $L$=60, 65, and 70 MeV.  The effective mass ratio for the nucleon, $(M_N^\ast/M_N)_0$, in SNM at $\rho_{\rm B}$ = $\rho_0$ is also listed. }~
\begin{center}
\begin{tabular}{ c || c | c | c | c | c || c | c || c }
\hline
 & $\gamma_a$ & $\eta_a$  &  $g_{\sigma N}$ & $g_{\omega N}$  & $g_{\rho N}$ & $\langle\sigma\rangle_0$ & $\langle\omega_0\rangle_0$  & $(M_N^\ast/M_N)_0$   \\ 
  & (MeV$\cdot$fm$^6$) & (fm$^3$) &          &         &  (MeV) &  (MeV) &  (MeV)  \\    \hline\hline
SJM2+TNA-L60    & $-$1662.63     &  17.18 &  5.27 & 8.16  & 3.29 & 39.06 & 16.37  & 0.78   \\
SJM2+TNA-L65    & $-$1597.67     &  18.25 &  5.71 & 9.07  & 3.35 & 42.16 & 18.18  & 0.74   \\
SJM2+TNA-L70    & $-$1585.48     &  19.82 &  6.07 & 9.77  & 3.41 & 44.62 & 19.59  & 0.71   \\
\hline
\end{tabular}
\label{tab:para}
\end{center}
\end{table*}

One finds that both the TNR and TNA have substantial contributions to the binding energy at $\rho_0$; $E$~(TNR)= 4.1 MeV and $E$~(TNA)=$-$6.6 MeV, which are similar to those of LP~(1981); $E$~(TNR)= 3.5 MeV and $E$~(TNA)=$-$6.1 MeV. It should be noted that the energy contribution from the three-$B$ force is determined by the {\it volume-integral} ${\cal V}$ of the effective two-body potential,
 i.~e.~${\cal V}$(UTBR) $\sim U_{\rm UTBR}(0;~\rho_{\rm B})\cdot \left(\lambda_r\right)^3$ in the case of UTBR with $U_{\rm UTBR}(0;~\rho_{\rm B})$ being the repulsive-core height and $\lambda_r$ the range of the potential, not by $U_{\rm UTBR}(0;~\rho_{\rm B})$ or $\lambda_r$ separately. 
 Actually, there is a large difference of both the height and the range of the potentials between SJM2 and LP~(1981) : $U_{\rm SJM2}(0;~\rho_0)$ =15.6 MeV and $\lambda_r$~(SJM2)=0.86 fm, while $U_{\rm TNR, LP}(0;~\rho_0)$= 2.89 MeV and $\lambda_r$~(LP) = 1.40 fm.
Nevertheless ${\cal V}$~(SJM2) coincides with ${\cal V}$~(LP) at $\rho_{\rm B}$=$\rho_0$, and both potentials contribute to almost the same amount of the $E$~(TNR) at $\rho_{\rm B}$=$\rho_0$. \\
\ \ One can see, in Fig.~\ref{fig1}, the difference of the $E$~(TNR) between SJM2 and LP(1981) is tiny (0.6 MeV) at $\rho_{\rm B}=\rho_0$, but it becomes large at high densities $\rho_{\rm B}\gtrsim$ 0.40~fm$^{-3}$ due to the sensitive density-dependence of the $E$~(TNR) ($\propto\rho_{\rm B}^2$). The stiffness of the EOS in the present model comes partially from such strong repulsion of the $E$~(TNR) at high densities. 
In addition, for two-body interaction in the RMF picture, attraction by the $\sigma$-meson exchange is saturated  at some density, while repulsion by the $\omega$-meson exchange increases steadily. This relativistic effect  also renders stiff EOS as compared with the nonrelativistic variational method of LP~(1981). \\
\ \ The value of $\gamma_a$ in $E$~(TNA) is correlated with a choice of the slope $L$ through the relation, 
$\Delta L$~(TNA)= $6\gamma_a\rho_0^2 (\eta_a\rho_0 - 2)e^{-\eta_a\rho_0}$ ($<$ 0), where $\Delta L$~(TNA) is the contribution to $L$ from the $E$~(TNA), stemming from the isospin-dependence of the $E$~(TNA).
Therefore a larger $L$ corresponds to a smaller $|\gamma_a|$, where the TNA has a less contribution to the binding energy at $\rho_{\rm B}=\rho_0$, $\Delta B_0$~(TNA). 
By adjusting to the $\Delta B_0$~(TNA), 
the coupling constants $g_{\sigma N}$, $g_{\omega N}$, $g_{\rho N}$ and meson mean-fields at $\rho_0$, $\langle\sigma\rangle_0$, $\langle\omega_0\rangle_0$ are modified to keep the saturation properties of the SNM. 
As a result, for a larger $L$, a contribution to the binding energy from the two-body $B$-$B$ interaction through the $\sigma$ and $\omega$ meson-exchange in the RMF framework gets larger, and so are the coupling constants and meson mean-fields at $\rho_0$ (see Table \ref{tab:para}). 
At high densities beyond $\rho_0$, where attraction from the $\sigma$-exchange is saturated, the remaining repulsion from the $\omega$-exchange is more marked for larger $\omega$-mean field, so that the stiffness of the EOS stands out as the $L$ increases from 60 MeV to 70 MeV. Hence, in our model, the slope $L$ controls the stiffness of the EOS in SNM not only around $\rho_0$, but also at high densities. This feature is applied also to hadronic matter with the ($Y$+$K$) phase. (see Sec.~\ref{sec:resultsKC} and Sec.~\ref{sec:MR}). 

As for the experimental constraints of the EOS of the SNM, the flow of matter in heavy ion collisions was analyzed to determine the pressures at density region $\rho_{\rm B}= (2 \sim 5) \rho_0$~\cite{danielewicz2002}. Our results in the present model show that the pressure-density curve in the SNM passes slightly above the upper limit of the allowable region constrained from the experimental data, which implies some softening may occur in the SNM for the relevant densities. 

\vspace{-0.6cm}~
\section{Onset density of KC and EOS for the ($Y$+$K$) phase}
\label{sec:resultsKC}

\begin{figure*}[h]~\vspace{-2.5cm}~
\begin{center}
\includegraphics[height=0.34\textheight]{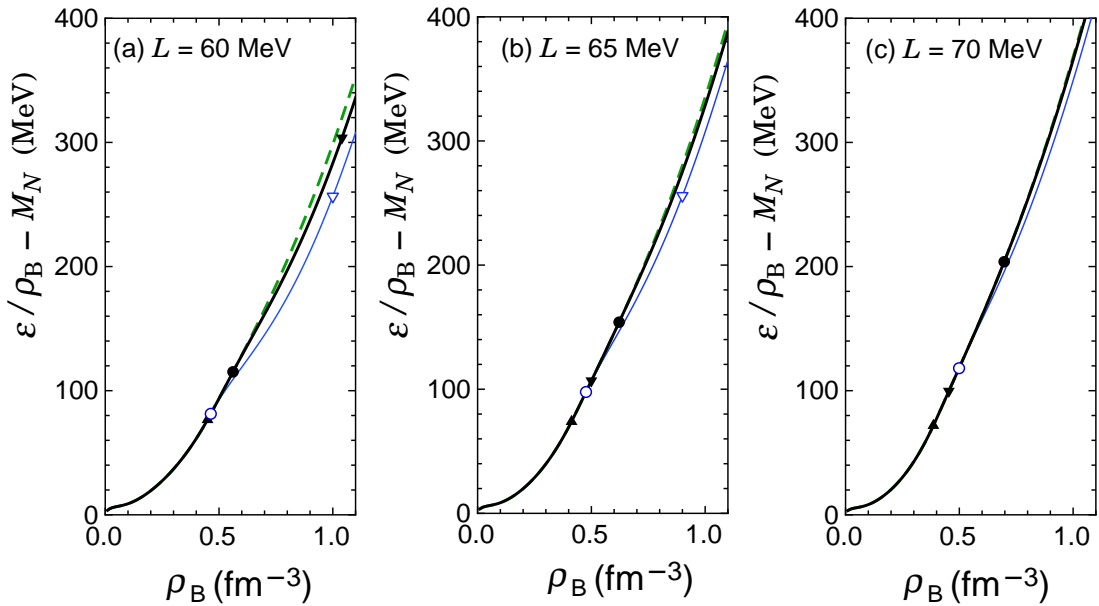}~
\end{center}~\vspace{1.5cm}~
\caption{The energy per baryon ${\cal E}/\rho_{\rm B}$ with the ($Y$+$K$) phase measured from the free nucleon mass as functions of baryon number density $\rho_{\rm B}$ for (a) $L$= 60 MeV, (b) $L$ = 65 MeV, and (c) $L$ = 70 MeV, obtained with the MRMF+UTBR+TNA. In each figure, the bold solid line is for $\Sigma_{Kn}$ = 300 MeV and the thin solid line is for $\Sigma_{Kn}$ = 400 MeV, respectively. 
For comparison, the energy per baryon for pure hyperon-mixed matter, where KC is switched off by setting $\theta=0$, is shown by the green dashed line. 
The onset densities of KC [$\rho_{\rm B}^c$ (K)], $\Lambda$ [$\rho_{\rm B}^c(\Lambda)$], and $\Xi^-$ [$\rho_{\rm B}^c(\Xi^-)$] in the case of $\Sigma_{Kn}$ = 300 MeV are denoted by the filled circle, filled triangle, and filled inverted triangle, respectively. The onset densities for KC and $\Xi^-$  in the case of $\Sigma_{Kn}$ = 400 MeV are denoted by the open circle and open inverted triangle, respectively. See the text for details.
}
\label{fig2}
\end{figure*}

In Fig.~\ref{fig2}, the energy per baryon ${\cal E}/\rho_{\rm B}$ with the ($Y$+$K$) phase measured from the free nucleon mass is shown as functions of baryon number density $\rho_{\rm B}$ for (a) $L$= 60 MeV, (b) $L$ = 65 MeV, and (c) $L$ = 70 MeV. In each figure, the bold solid line is for $\Sigma_{Kn}$ = 300 MeV and the thin solid line is for $\Sigma_{Kn}$ = 400 MeV, respectively. 
For comparison, the energy per baryon for pure hyperon-mixed matter, where KC is switched off by setting $\theta=0$, is shown by the green dashed line. 
The onset densities of KC [$\rho_{\rm B}^c~(K)$], $\Lambda$ [$\rho_{\rm B}^c(\Lambda)$], and $\Xi^-$ [$\rho_{\rm B}^c(\Xi^-)$] in the case of $\Sigma_{Kn}$ = 300 MeV are denoted by the filled circle, filled triangle, and filled inverted triangle, respectively. The onset densities for KC and $\Xi^-$  in the case of $\Sigma_{Kn}$ = 400 MeV are denoted by the open circle and open inverted triangle, respectively. 
Note that the $\Lambda$ hyperons always precede KC for (a), (b), (c), so that $\rho_{\rm B}^c(\Lambda)$ is common to both cases of $\Sigma_{Kn}$ = 300 MeV and 400 MeV. 
For $L$ = 65 MeV and 70 MeV with $\Sigma_{Kn}$ = 300 MeV, even $\Xi^-$ hyperons appear at lower density than KC. 
Mixing of the negatively charged hyperons $\Xi^-$ pushes the onset of KC to high densities, so that $\rho_{\rm B}^c(K)$ is delayed to higher densities for larger $L$. 
Conversely, in case KC onsets at a lower density than the $\Xi^-$ hyperons (for $L$ = 60 MeV with $\Sigma_{Kn}$ = 300 MeV and for $L$ = (60, 65, 70) MeV with $\Sigma_{Kn}$ = 400 MeV), mixing of the $\Xi^-$ hyperons is pushed up to high densities, or even does not occur over the relevant densities: 
KC and $\Xi^-$ hyperons compete against each other through the repulsive $K$-$\Xi^-$ vector interaction term in $X_0$, the form of which is dictated by chiral symmetry. 

From Fig.~\ref{fig2}, the onset density of KC is read as 
$\rho_{\rm B}^c(K)$ = (0.56, 0.62, 0.70)~fm$^{-3}$ [ (0.46, 0.48, 0.50)~fm$^{-3}$] for $L$ = (60, 65, 70) MeV in the case of $\Sigma_{Kn}$ = 300 MeV (400 MeV). 
The appearance of KC in the hyperon ($\Lambda$)-mixed matter 
leads to further decrease in energy of the system due to the $s$-wave $K$-$B$ attraction
from that due to the $\Lambda$-mixing. As a result, 
the EOS for the ($Y$+$K$) phase is further softened in comparison with that in the pure $Y$-mixed matter (the dashed lines in Fig.~\ref{fig2}). There is a clear difference in energy for $\Sigma_{Kn}$ = 400 MeV from the case of the pure $Y$-mixed matter, while 
the difference is tiny for $\Sigma_{Kn}$ = 300 MeV, in particular, in the case of $L$ = 70 MeV. 
It should be noted that the stronger three-baryon repulsion as a net effect of the UTBR and TNA leads to more remarkable saturation and subsequent reduction of the nuclear scalar density $\rho_{p,n}^s$ as a relativistic effect. As a result, a part of the attractive energy from KC, which comes from the effective baryon mass term in (\ref{eq:lagbm}) being proportional to $\rho_N^s\Sigma_{KN}$, is suppressed more, so that the decrease in energy due to KC is moderated by the introduction of the three-baryon force, leading to suppression of the significant softening of the EOS even in the presence of KC. 

\section{Mass-radius relations of kaon-condensed neutron stars}
\label{sec:MR}

\begin{figure}[h]~\vspace{-2.5cm}~\hspace{0.5cm}~
\begin{center}
\includegraphics[height=.40\textwidth]{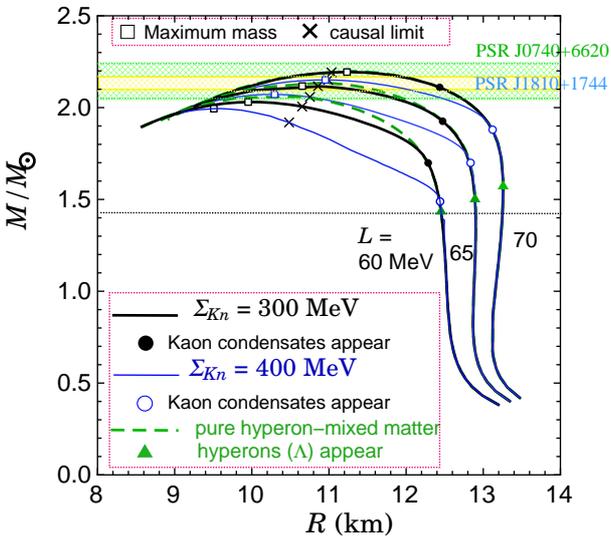}~
\end{center}~\vspace{1.2cm}~
\caption{The gravitational mass~$M$ to radius~$R$ relations after solving the Tolman-Oppenheimer-Volkoff  equation for $L$ = (60, 65, 70) MeV obtained with MRMF+UTBR(SJM2)+TNA. The branches including KC in the core are denoted as the black bold solid lines (blue thin solid lines) for $\Sigma_{Kn}$ = 300 MeV (400 MeV).  For comparison, the branch with pure $Y$-mixed matter, where KC is switched off by setting $\theta=0$, is shown by the green dashed line for each case of $L$. }
\label{fig3}
\end{figure}

Based on the EOS including the ($Y$+$K$) phase, we discuss the effects of KC on the structure of compact stars. 
In Fig.~\ref{fig3}, the gravitational mass $M$ - radius $R$ relations after solving the Tolman-Oppenheimer-Volkoff equation are shown for $L$ = (60, 65, 70) MeV, obtained with the MRMF+UTBR+TNA. The branches including KC in the core are denoted as the black bold solid lines (blue thin solid lines) for $\Sigma_{Kn}$ = 300 MeV (400 MeV).  For comparison, the branch including pure hyperon-mixed matter, where KC is switched off by setting $\theta=0$, is shown by the green dashed line for each case of $L$. The green filled triangle [$\blacktriangle$] stands for the branch point where the $\Lambda$ hyperons appear from nuclear matter in the center of the star. The branch point at which KC appears in the center of the star is indicated by the filled circle [$\bullet$] (open circle [$\circ$]) in the case of $\Sigma_{Kn}$ = 300 MeV (400 MeV). 
The maximum mass point for each branch including the ($Y$+$K$) phase is indicated by the open square [$\square$]. The cross point [$\times$] corresponds to the causal limit at which the sound velocity exceeds the speed of light. 
One can see, in Fig.~\ref{fig2}, that the EOS becomes stiffer at high densities for larger $L$. This fact is reflected in that the maximum mass and its radius shift to larger values for larger $L$, as seen in Fig.~\ref{fig3}. Also the radius of neutron stars for a given mass in the stable branch increases with $L$. \\
\ \ The maximum masses for $L$ =(65, 70) MeV are consistent with recent observations of massive neutron stars in both cases of $\Sigma_{Kn}$ = 300 MeV and 400 MeV, while the masses within the causal limit for $L$ = 60 MeV do not reach the range allowable from the observations of most massive neutron stars to date~\cite{c2020,romani2021} (the green and yellow bands in Fig.~\ref{fig3}). The radius $R$ in the stable branches is consistent with observational constraints from gravitational waves of the binary neutron star mergers GW170817~\cite{abbott2018,horowitz2018}. Also $R$ for $M\simeq 1.3M_\odot$ lie within the range of the mass and radius deduced from NICER observations of PSR~J0030+0451~\cite{riley2019,miller2019}. 
The observation of a compact object with a mass of (2.50$-$2.67) $M_\odot$ in the GW190814 provoked a question if the second companion is the heaviest neutron star or the lightest black hole~\cite{abbott2020}. The former case will provide the stringent constraint on the EOS of dense matter, although further detailed analyses are needed to obtain a definite conclusion. 

 It is expected that heavy neutron stars with $M \gtrsim 1.7~M_\odot$ have the ($Y$+$K$) core. A large portion of the core may be occupied with ($Y$+$K$) phase for massive neutron stars: 
For neutron stars with $M$ = 2.0 $M_\odot$  in the case of $L$ = 65 MeV and $\Sigma_{Kn}$ = 300 MeV  ($L$ = 70 MeV and  $\Sigma_{Kn}$ = 400 MeV), one has the ($Y$+$K$) core composed of KC, $\Lambda$ and $\Xi^-$-mixed baryonic matter within the region of radius 3.8~km (4.8~km).  
On the other hand, for neutron stars with $M\lesssim$1.4 $M_\odot$, the central density does not reach $\rho_{\rm B}^c (\Lambda)\sim 0.4$~fm$^{-3}$, and the ground state in the core consists of only $n$, $p$, and leptons ($e^-$). 

Recently, it has been pointed out that the $\Delta^-$ isobars may be mixed at densities similar to those of hyperon-mixing in neutron stars for the values of the slope, $L=(40-60)$ MeV~\cite{d2014}. 
However, there is large ambiguity about the empirical $\Delta$-meson coupling constants and the $\Delta$ isobar potential in matter~\cite{kmv2017}. For the sake of brevity, we don't consider possibility of mixing of the $\Delta^-$, and we concentrate on making clear the suppression mechanisms of KC in the hyperon-mixed matter.

\section{Summary and concluding remarks}
\label{sec:summary}

We have shown that the ($Y$+$K$) phase can be realized in neutron stars with $M\gtrsim$ 1.7 $M_\odot$, depending on the allowable values of $L$ [= (60$-$70) MeV] and $\Sigma_{Kn}$ [=(300$-$400) MeV]. The EOS and the resulting mass and/or radius of compact stars within hadronic picture accompanying the ($Y$+$K$) phase are consistent with recent observations of massive neutron stars. 

In this work, we have fixed the UTBR to the SJM2 model. We should consider how systematic relaxing of the volume integral for the UTBR affects the stiffness of the EOS including the ($Y$+$K$) phase. 
Validity of the UTBR should also be examined by comparing with other results in quark models including the quark Pauli effects~\cite{oka2012,nakamoto2016} and lattice QCD results~\cite{inoue2019}. 
 
In the presence of KC, rapid cooling mechanisms through $\nu$, $\bar\nu$ emissions may be kinematically possible, which plays an important role on thermal evolution of neutron stars:  One is the kaon-induced Urca (KU) process, $N + \langle K^-\rangle\rightarrow N + e^- +\bar\nu_e$, $N + e^- \rightarrow N + \langle K^- \rangle + \nu_e$ ($N=p,n$), where $\langle K^- \rangle$ stands for the classical $K^-$ field which supplies the system with energy $\mu_K$ to make the reaction kinematically possible~\cite{t88,bkpp88}. The other is the direct Urca (DU) process in KC, $n\rightarrow p+e^- +\bar\nu_e$, $p+e^-\rightarrow  n + \nu_e$, as long as the kinematical condition for the reaction is met depending upon the density-dependence of the symmetry energy~\cite{fmtt94}. 
According to the results in the present work,  
main cooling process is divided by the mass $M \sim1.4 M_\odot$. For $M\lesssim 1.4 M_\odot$, it is given by the modified Urca process since the proton-mixing ratio is under threshold for the DU process, $\rho_p/\rho_{\rm B}\lesssim1/9$.
For $M > 1.4 M_\odot$, Hyperon ($\Lambda$) Urca process, $\Lambda\rightarrow p+e^- +\bar\nu_e$, $p+e^-\rightarrow  \Lambda + \nu_e$~\cite{plp1992} starts and becomes a dominant cooling process. For  the massive neutron stars ($M \gtrsim 1.7~M_\odot$), the KU process becomes a main cooling process. 
Several neutron stars have anomalously low temperature that requires extraordinary rapid cooling processes~\cite{tsuruta1998,doi2019}. Unified description of emissivities for these reactions with composition of matter is indispensable.
 
Two of the authors (T.~Maruyama and T.~Tatsumi) considered a pasta structure of kaon condensed phase~\cite{mtv2006}. According to the previous result without three-baryon forces, significant softening accompanying the KC pasta lead to the transition of first order. In the present result with the three-baryon forces,  the transition to KC becomes of second order, which may modify various aspects of the KC pasta structure. 

Throughout this work, we have concentrated on the $s$-wave KC for simplicity. In the presence of hyperons, the $p$-wave $KNY$ interaction necessarily arises in addition to the $s$-wave $K$-$B$ interaction. It has been shown that a spontaneous creation of a pair of the particle-hole collective modes with $K^+$ and $K^-$ quantum numbers ($p$-wave kaon condensation) may occur at densities where $\Lambda$ hyperons are more abundant than protons~\cite{muto2002}. The three-baryon forces 
may affect not only the onset and EOS of the $p$-wave kaon condensed phase but also rapid cooling mechanisms associated with the $p$-wave kaon condensation. 
 
As another picture for stiffening the EOS, strange quark matter and hadron-quark phase transition have been studied extensively. In particular, hadronic matter was connected to quark matter smoothly by the crossover transition to obtain massive neutron stars compatible with observations~\cite{mht2013,baym2018}. The connection of hadronic phase with quark degrees of freedom at high densities will be considered in future works.  

 \vspace{-0.5cm}~
\section*{Acknowledgement}
 One of the authors (T.~Muto) acknowledges the financial support by Chiba Institute of Technology. 
We thank the late Prof. T.~Takatsuka for his collaboration on the present subject. 

\vspace{-0.7cm}~

\end{document}